\documentclass[showpacs,amsmath,amssymb,aps,twocolumn]{revtex4-1}
\usepackage{graphics}
\usepackage{graphicx}
\usepackage{txfonts}
\usepackage{bm}   
\usepackage{amsmath,mathrsfs,amssymb}
\usepackage{multirow}
\begin{document}

\title{Global study of nuclear modifications on parton distribution functions}

\author{Rong Wang$^{1,2,3}$}\email{email: rwang@impcas.ac.cn}
\author{Xurong Chen$^{1}$}\email{email: xchen@impcas.ac.cn}
\author{Qiang Fu$^{1,2,3}$}

\affiliation{
$^1$ Institute of Modern Physics, Chinese Academy of Sciences, Lanzhou 730000, China\\
$^2$ Lanzhou University, Lanzhou 730000, China\\
$^3$ University of Chinese Academy of Sciences, Beijing 100049, China\\
}

\date{\today}

\begin{abstract}
A global analysis of nuclear medium modifications of parton
distributions is presented using deeply inelastic scattering data
of various nuclear targets. Two obtained data sets are provided
for quark and gluon nuclear modification factors, referred as nIMParton16.
One is from the global fit only to the experimental data of
isospin-scalar nuclei (Set A), and the other is from the fit to
all the measured nuclear data (Set B). The scale-dependence is
described by DGLAP equations with nonlinear corrections in this work.
The Fermi motion and off-shell effect, nucleon swelling, and parton-parton
recombination are taken into account together for modeling the complicated
$x$-dependence of nuclear modification. The nuclear gluon shadowing
in this paper is dynamically generated by the QCD evolution of parton
splitting and recombination processes with zero gluon density at the
input scale. Sophisticated nuclear dependence of nuclear medium effects
is studied with only two free parameters. With the obtained free parameters
from the global analysis, the nuclear modifications of parton distribution
functions of unmeasured nuclei can be predicted in our model.
Nuclear modification of deuteron is also predicted and shown with
recent measurement at JLab.
\end{abstract}

\maketitle

\section{Introduction}
\label{SecI}

Parton distribution functions (PDFs) of nuclei are different from
PDFs of free nucleon, which clearly indicates that the quark
or gluon freedom inside the bound nucleon is influenced by the nuclear
medium environment. The nuclear medium effect on quark or gluon
distribution attracts a lot of interests on both the experimental
and theoretical sides, particularly since the discovery of
the EMC effect in the valence-dominant region\cite{emc-discovery}.
The nuclear medium modifications on PDFs are usually depicted by
the nuclear modification factor defined as the per-nucleon structure
function ratio $R=F_2^A/F_2^{D}$, since deuteron is approximately viewed
as a system of free proton and free neutron. The nuclear modifications
are commonly categorized as the shadowing effect, anti-shadowing effect,
the EMC effect, and Fermi motion smearing according to the different shapes
of the ratio $R$ in different $x$ regions\cite{review1,review2,review3,review4}.
The shadowing effect refers
to $R<1$ in $x\lesssim 0.1$ region; The anti-shadowing effect refers to
a small enhancement of $R>1$ in range of $0.1\lesssim x \lesssim 0.3$;
The EMC effect refers to the slope of $R$ in the valence-dominant region
$0.3\lesssim x \lesssim 0.7$; The Fermi motion refers to the rising of $R$
in the range of $x\gtrsim 0.7$.
The $x$-dependence and nuclear dependence of nuclear modification are
so complicated that there is no commonly accepted theory which explains
the nuclear corrections in whole $x$ range for all nuclei so far.
It is commonly acknowledged that different mechanisms are responsible
for the different nuclear modifications in different regions of $x$.

Nuclear medium modification is a fundamental issue for high energy nuclear
physics, because the accurate nuclear PDFs (nPDFs) of various nuclei are
indispensable for simulations/calculations of relativistic nucleus-nucleus
reactions on modern ion colliders - RHIC at BNL and LHC at CERN\cite{RHIC,LHC-1,LHC-2}.
On the one hand, nPDFs provide the pre-collision condition of colliding nuclei,
which plays an important role in quantifying the microscopic processes of
the evolution of QCD matter during relativistic heavy-ion collision.
It is also vital in the study of vector boson (like $J/\Psi$ and $\Upsilon$), jets,
Z-boson, or other hadron production in nucleus-nucleus collisions.
On the other hand, the precise and detailed nuclear corrections
allow us to combine data across different nuclear targets and provide
maximum information for the proton PDFs, such as the difference
between up and down sea quark distributions. Massive targets are usually used
to get sufficient statistics for neutrino-nucleus deeply inelastic
scattering (DIS) measurements, as the neutrino cross section is very small.
The neutrino-nucleus DIS data is important for the separation of different
flavor components of PDFs. Therefore, precise nuclear modifications are required
if we want to include the neutrino DIS data into the global analysis of proton PDFs.

There are a lot of global analyses of nPDFs worked out so far,
such as EKS\cite{EKS98}, EPS\cite{EPS08,EPS09}, nCTEQ\cite{nCTEQ09,nCTEQ10,nCTEQ15},
nDS\cite{nDS}, DSSZ\cite{DSSZ}, HKN\cite{HKN04,HKN07}, KP\cite{KP06,KP07,KP14,KP-DY},
and KT\cite{KT16}, showing some remarkable progresses.
All of these analyses were performed with
the application of DGLAP equations, and the collinear factorization also works
good for the nuclear PDFs case. The main difference among the widely used nPDFs
is the technique of parameterizing the nuclear modifications at
the initial scale of $Q_0^2>1$ GeV$^2$. For EKS and EPS\cite{EKS98,EPS08,EPS09},
nuclear modification factors of valence quark ($R_V$), sea quark ($R_S$)
and gluon distributions ($R_G$) are parameterized at the input scale, under
the assumptions of $R_{uv}=R_{dv}=R_V$ and $R_{\bar{u}}=R_{\bar{d}}=R_{\bar{s}}=R_S$.
For the early analysis of EKS98, $R_G=R_{F_2}$ is assumed in order to
get a stable evolution of nPDFs\cite{EKS98}. For nCTEQ analyses, the process
is straight forward with the nuclear initial parton distributions of $u_v$, $d_v$,
$g$, $\bar{u}+\bar{d}$, $s$ and $\bar{s}$ parameterized instead of the nuclear
modification factors\cite{nCTEQ09,nCTEQ10,nCTEQ15}. The technique of nCTEQ analysis
is in close analogy to the proton global analysis. For nDS global analysis,
the nuclear modification is parameterized by using a convolution method\cite{nDS}
with three different convolution weights for valence quark, sea quark
and gluon distributions respectively. The nuclear modification parameterization
of DSSZ is similar to that of EKS except that the modification factors
$R_V$, $R_S$ and $R_G$ are not independent\cite{DSSZ}. For HKN analyses,
the modification factors of $u_v$, $d_v$, $\bar{q}$ and $g$ are used.
It is shown in works of HKN that the nuclear gluon distributions cannot be fixed
by the present data\cite{HKN04,HKN07}.
KP analyses are very successful for description of the nuclear structure
functions in the entire kinematical region of $x$ and $Q^2$,
by developing a model which takes into account of nuclear shadowing,
Fermi motion and binding, nuclear pion excess and off-shell correction\cite{KP06,KP07}.
The semimicroscopic model by KP also gives nPDFs information using impulse approximation
with corrections of coherent nuclear interactions and of nuclear
meson-exchange currents\cite{KP14,KP-DY}, which has a successful application
in proton-lead collisions\cite{RKPZ}.
KT analysis of nPDFs\cite{KT16} is the global analysis performed
at next-to-next-to-leading order for the first time.

Gluon distribution, nuclear dependence and spatial (impact parameter) dependence
of nuclear effect are three main problems needs to be address for further improving
the global analysis of nPDFs\cite{HKN04,HKN07,Eskola-rev}. Due to the challenge
of not enough experimental data, limited kinematic coverage of the data points
and many free parameters ($>15$), gluon distribution can not be fixed and
it has large uncertainty in the whole $x$ region\cite{HKN04,HKN07}.
Valence quark distribution at small $x$ and anti-quark distribution at large $x$
in nuclei are also not clear. Accurate nuclear gluon distribution can be obtained
if reliable constraints of PDFs are applied. The dynamical parton model
has a good way to constrain the gluon distribution. The gluon distribution is zero
at extremely low input scale\cite{IMParton,Chen-proton}, where gluon distribution
at high $Q^2$ is dynamically generated through the parton splitting and parton-parton
recombination processes during QCD evolution. Our previous works\cite{IMParton,Chen-proton}
show that the dynamical parton model agrees well with the experimental data.
The strength of parton-parton recombination or the correlation length $\tilde{R}$ for
two-parton distribution is already fixed in the global analysis
of proton PDFs\cite{IMParton}. The only question left is the recombination enhancement
due to the recombination of partons of small $x$ between two different nucleons in a nucleus.
On the process of two-parton recombination, the enhancement is proportional to
the size of nucleus ($A^{1/3}$). To constrain nuclear gluon distribution
under the dynamical parton model is the main purpose of this work.

PDFs are of nonperturbative origin. Our previous work\cite{IMParton} has obtained
a nonperturbative input with only three valence quarks at $Q_0^2=0.0671$ GeV$^2$.
This provides us a good opportunity to apply the nonperturbative nuclear effects
(nucleon swelling, binding effect and Fermi motion) on the nonperturbative input.
Therefore, two questions are encountered for us to get nPDFs. One is how the nuclear
effects modify the initial nPDFs, and the other is how the nuclear medium effects depend
on the nuclear targets. This analysis is based on
some models with clear mechanisms, which gives a way to know how the nuclear
effects modify the nonperturbative input. In this work, the $x$-dependence of nuclear
modification factor in whole $x$ range is attributed to the composited influence
of parton-parton recombination corrections, nucleon swelling, binding effect
and Fermi motion. The last decade has seen a lot of progresses on the nuclear dependence
of the EMC effect\cite{NuclDepenRev,RSIE}. Local nuclear density is thought
to be the most relevant quantity to explain the magnitude of the EMC effect.
A novel nuclear dependence of the EMC effect\cite{RSIE} is used in this paper.

This article is organized as follows. The charged lepton-nucleus DIS data we used
are listed in Sec. \ref{SecII}. The convolution method of Fermi smearing
and the binding effect are shown in Sec. \ref{SecIII}. The nucleon swelling effect on
the standard deviation of valence quark distributions is discussed in Sec. \ref{SecIV}.
DGLAP equations with parton-parton recombination corrections are shown in Sec. \ref{SecV},
which are used to describe the $Q^2$-dependence of nPDFs. In Sec. \ref{SecVI},
procedure of the global analysis is presented. In Sec. \ref{SecVII}, we show
the global fit results, comparisons of the obtained nPDFs with experimental data
and with other widely used nPDFs. In Sec. \ref{SecVIII}, a C++ program is introduced
for obtaining the nuclear modification factors of many nuclei predicted by this work.
At last, some discussions and a brief summary are given in Sec. \ref{SecIX}.

\section{Experimental data}
\label{SecII}

\begin{table}[htp]
\centering
\caption{
List of nuclear data used in the analysis.
Number of data points after kinematic cuts are shown.
The individual values of $\chi^2$ of the global fits are also shown
for each experimental set (see Sec. \ref{SecVII}).
}
\begin{tabular}{cccccc}
\hline
ID   &  Targets  & Experiment & \# of data &~~$\chi^2$ of Fit A~~&~~$\chi^2$ of Fit B~~\\
\hline
1    & $^4$He/$^2$H      & NMC95\cite{NMC441.3}  & 15   & 12.57 & 15.01 \\
2    & $^4$He/$^2$H      & SLAC\cite{SLAC94}     & 18   & 25.74 & 47.39 \\
3    & $^4$He/$^2$H      & JLab\cite{JLab09}     & 11   & 11.41 & 20.47 \\
4    & $^6$Li/$^2$H      & NMC\cite{NMC441.12}   & 14   & 17.14 & 16.05 \\
5    & $^{12}$C/$^2$H    & NMC\cite{NMC441.3}    & 15   & 20.07 & 23.72 \\
6    & $^{12}$C/$^2$H    & NMC\cite{NMC441.12}   & 15   & 13.71 & 12.00 \\
7    & $^{12}$C/$^2$H    & SLAC\cite{SLAC94}     & 7    & 10.62 & 2.644 \\
8    & $^{12}$C/$^2$H    & JLab\cite{JLab09}     & 27   & 33.54 & 57.23 \\
9    & $^{12}$C/$^2$H    & EMC\cite{EMC202.603}  & 9    & 7.505 & 7.938 \\
10    & $^{12}$C/$^2$H   & EMC\cite{EMC211.493}  & 8    & 22.37 & 23.07 \\
11    & $^{12}$C/$^2$H   & EMC\cite{EMC333.1}    & 6    & 20.45 & 20.95 \\
12    & $^{12}$C/$^2$H   & E665\cite{E66595}     & 4    & 10.46 & 8.243 \\
13    & $^{14}$N/$^2$H     & BCDMS\cite{BCDMS85} &  9   & 9.198 & 6.808  \\
14    & $^{40}$Ca/$^2$H  & NMC\cite{NMC441.3}    & 14   & 20.00 & 21.46 \\
15    & $^{40}$Ca/$^2$H  & SLAC\cite{SLAC94}     & 7    & 6.103 & 11.76 \\
16    & $^{40}$Ca/$^2$H  & EMC\cite{EMC211.493}  & 8    & 12.06 & 14.36 \\
17    & $^{40}$Ca/$^2$H  & EMC\cite{EMC333.1}    & 6    & 15.55 & 16.44 \\
18    & $^{40}$Ca/$^2$H  & E665\cite{E66595}     & 4    & 12.70 & 7.977 \\
19    & $^{12}$C/$^6$Li    & NMC\cite{NMC53.73}  & 19   & 30.34 & 31.69 \\
20    & $^{12}$C/$^6$Li    & NMC\cite{NMC441.3}  & 19   & 30.06 & 31.44 \\
21    & $^{40}$Ca/$^6$Li   & NMC\cite{NMC53.73}  & 19   & 22.29 & 26.72 \\
22    & $^{40}$Ca/$^6$Li   & NMC\cite{NMC441.3}  & 19   & 21.85 & 25.09 \\
23    & $^{40}$Ca/$^{12}$C & NMC\cite{NMC53.73}  & 19   & 20.72 & 19.71 \\
24    & $^{40}$Ca/$^{12}$C & NMC\cite{NMC441.3}  & 19   & 21.70 & 19.51 \\
25    & $^{40}$Ca/$^{12}$C & NMC\cite{NMC481.3}  & 15   & 6.659 & 7.188 \\
\multicolumn{3}{c}{Sum of isospin-scalar nuclei} & 326  & 434.8 & 494.9 \\
26    & $^3$He/$^2$H       & JLab\cite{JLab09}   &  11  &   -   & 38.68  \\
27    & $^9$Be/$^2$H       & JLab\cite{JLab09}   &  11  &   -   & 31.25  \\
28    & $^9$Be/$^2$H       & SLAC\cite{SLAC94}   &  17  &   -   & 18.68  \\
29    & $^{27}$Al/$^2$H    & SLAC\cite{SLAC94}   &  17  &   -   & 17.45  \\
30    & $^{56}$Fe/$^2$H    & SLAC\cite{SLAC94}   &  23  &   -   & 44.27  \\
31    & $^{56}$Fe/$^2$H    & BCDMS\cite{BCDMS85} &  6   &   -   & 4.229  \\
32    & $^{56}$Fe/$^2$H    & BCDMS\cite{BCDMS87} &  10  &   -   & 20.89  \\
33    & Cu/$^{2}$H         & EMC\cite{EMC57.211} &  19  &   -   & 11.41  \\
34    & Ag/$^2$H           & SLAC\cite{SLAC94}   &  7   &   -   & 11.25  \\
35    & Xe/$^{2}$H         & E665\cite{E66592}   &  3   &   -   & 1.319  \\
36    & $^{197}$Au/$^2$H   & SLAC\cite{SLAC94}   &  18  &   -   & 94.65  \\
37    & $^{208}$Pb/$^{2}$H  & E665\cite{E66595}  &  15  &   -   & 14.64  \\
38    & $^{9}$Be/$^{12}$C   & NMC\cite{NMC481.3}  &  15 &   -   & 5.694  \\
39    & $^{27}$Al/$^{12}$C  & NMC\cite{NMC481.3}  &  15 &   -   & 6.650  \\
40    & $^{56}$Fe/$^{12}$C  & NMC\cite{NMC481.3}  &  15 &   -   & 9.366  \\
41    & Sn/$^{12}$C         & NMC\cite{NMC481.3}  &  15 &   -   & 17.85  \\
42    & Sn/$^{12}$C         & NMC\cite{NMC481.23} &  154 &   -  & 112.4  \\
43    & $^{208}$Pb/$^{12}$C & NMC\cite{NMC481.3}  &  15  &   -  & 10.65  \\
\multicolumn{3}{c}{Sum of all nuclear data}       & 701  &   -  & 966.2 \\
\hline
\end{tabular}
\label{table_data}
\end{table}

The charged lepton-nucleus deeply inelastic scattering has been
a powerful tool to study the nuclear structure and the nucleon structure for decades.
The high energy lepton probe is so clean that we only include the DIS data
in the analysis as a baseline for further studies.
The nuclear modification data of per-nucleon structure functions
or differential cross sections used in this work are taken from
EMC\cite{EMC202.603,EMC211.493,EMC333.1,EMC57.211},
NMC\cite{NMC53.73,NMC441.3,NMC441.12,NMC481.3,NMC481.23},
SLAC\cite{SLAC94}, BCDMS\cite{BCDMS85,BCDMS87},
Fermilab E665\cite{E66592,E66595}, and JLab\cite{JLab09} experiments.
Some kinematic cuts on the experimental data are used to make sure
the data are in the deep inelastic region, which is shown in Eq. (\ref{Q2W2_cuts}).
\begin{equation}
Q^2\ge 2~\text{GeV}^2, W^2\ge 4~\text{GeV}^2.
\label{Q2W2_cuts}
\end{equation}
After the selection by these kinematic cuts, we get 56, 431, 114, 25, 15 and
60 data points from EMC, NMC, SLAC, BCDMS, Fermilab E665 and JLab respectively.

\begin{figure}[htp]
\begin{center}
\includegraphics[width=0.41\textwidth]{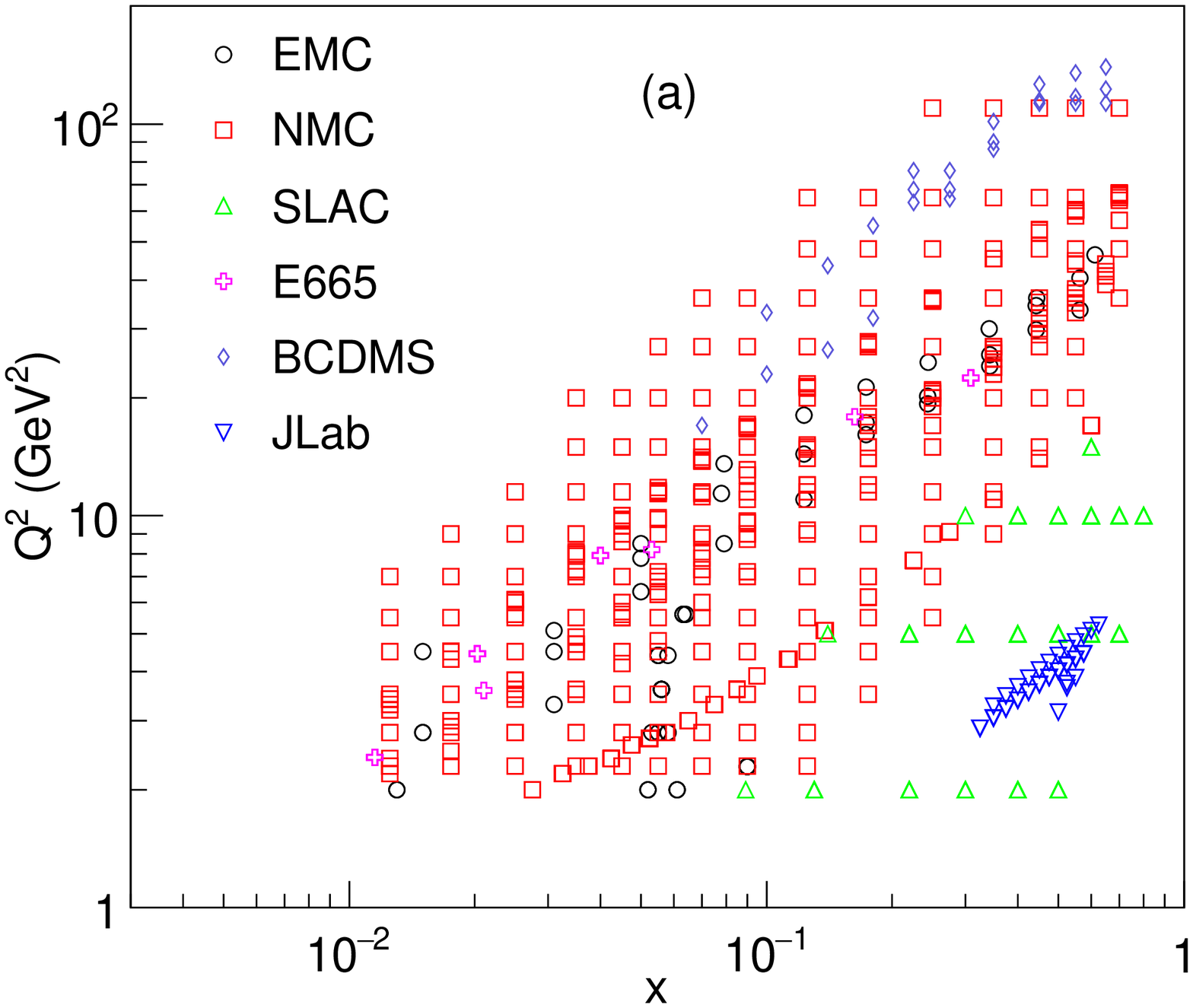}
\includegraphics[width=0.41\textwidth]{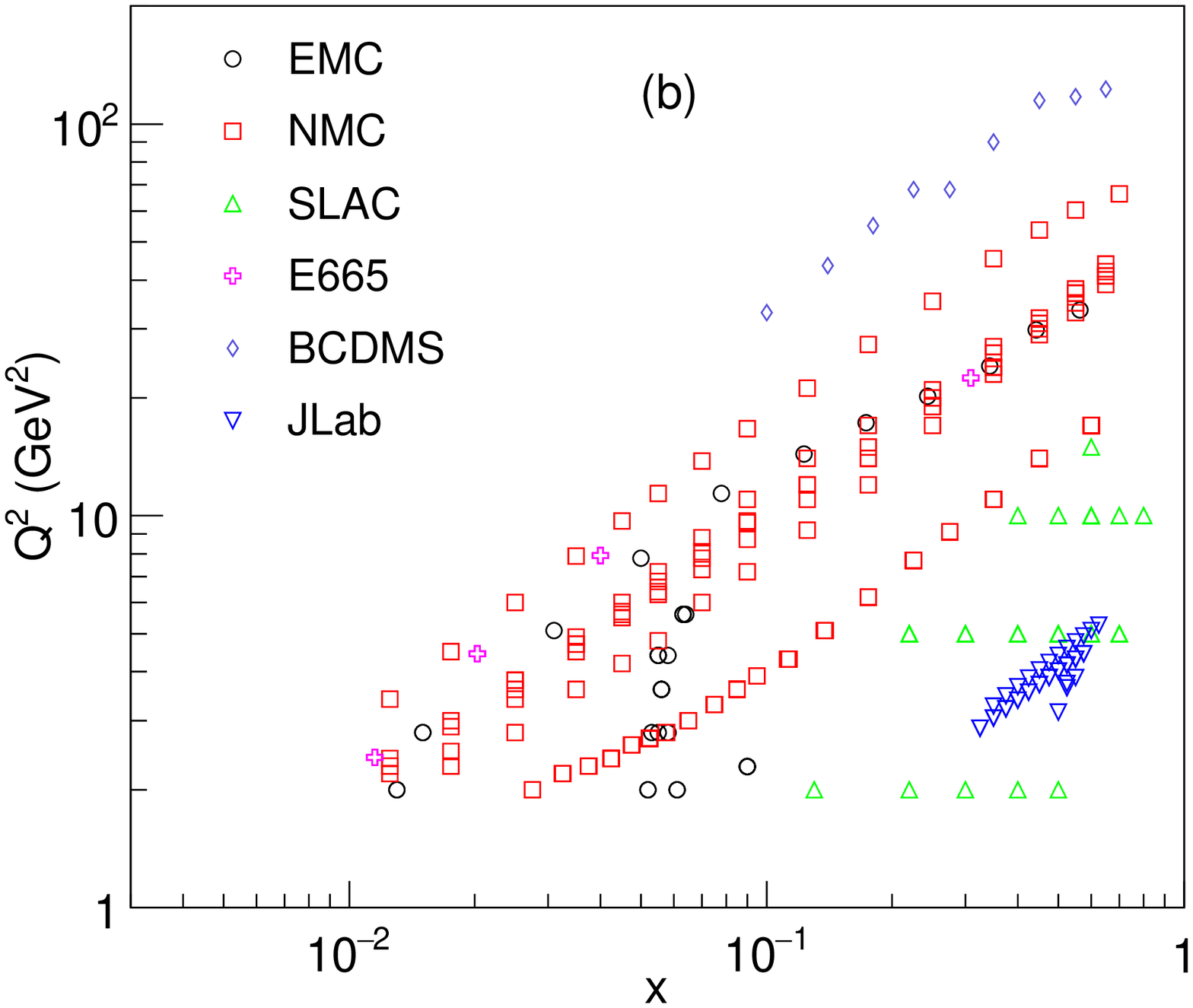}
\caption{
(a) The kinematic coverage of the worldwide nuclear DIS data for the global QCD analysis;
(b) The kinematic coverage of the nuclear DIS data of isospin-scalar nuclei.
}
\label{ExpKine}
\end{center}
\end{figure}

The measured nuclear targets used in the global analysis are listed in Table \ref{table_data}.
The kinematic ranges of nuclear data are shown in Fig. \ref{ExpKine} (a)
and Fig. \ref{ExpKine} (b) for isospin-scalar data and all nuclear data respectively.
Since the $Q^2$ of nuclear data are not high, the target mass correction
is performed to the DIS data. The formula of target mass correction
is taken from Refs. \cite{TMC-1,TMC-2}.

\section{Fermi motion and off-shellness}
\label{SecIII}

Compared to free nucleon, the nucleon in heavy nuclear target is moving
with average Fermi momentum and of off-shell kinematic.
For nuclear experimental data, the $x$ variable is determined
in the approximation that the nucleon is at rest.
Therefore the measured nuclear structure function is the convolution
of the bare nucleon structure function with the momentum distribution
function of a nucleon inside the nucleus.
Generally the Fermi smearing and binding effect deform
the nuclear structure functions mainly at large $x\gtrsim 0.7$.
In principle, nuclear parton distributions are non-zero in the range of $1<x<A$,
and Fermi smearing effect still exists in the region.
However, nPDFs in the tail region of $x>1$ are very small, which can be
neglected in most cases. Therefore, we only focus on nPDFs in range of
$0<x<1$ in this analysis.

In this work, the Fermi motion and binding effect are
taken into account to interpret the nuclear modification at large $x$.
These effects were well studied by Bodek and Ritchie\cite{Bodek1981},
and Frankfurt and Strikman\cite{Frankfurt1980}.
The deformation of the initial parton distributions is given by
the convolution method\cite{review1,Bodek1981,Frankfurt1980,Chen-EMC},
\begin{equation}
\begin{aligned}
xf_i^A(x,Q_0^2)=\int_{x}^Adyf_N(y)(x/y)f_i^N(x/y,Q_0^2),
\end{aligned}
\label{FermiConvol}
\end{equation}
where
\begin{equation}
\begin{aligned}
f_N(y)=
\begin{cases}
\frac{3m_N}{4k_F^3}[k_F^2-m_N^2(y-\eta_A)^2],~(\eta_A-k_F/m_N<y<\eta_A+k_F/m_N)\\
0,~(y<\eta_A-k_F/m_N~or~y>\eta_A+k_F/m_N)\\
\end{cases}
\end{aligned}
\label{NuclWaveFunc}
\end{equation}
with $k_F$ the average Fermi momentum, $m_N$ the nucleon mass,
and $\eta_A=1-B_A/m_N$, in which the $B_A$ is
the nuclear binding energy per nucleon.
The average nucleon Fermi momentum for the nuclei is around 200 MeV/c,
and the binding energy is taken from Ref. \cite{MengWang}, which is precisely measured.

The last important thing to evaluate the Fermi motion effect is to estimate
the average Fermi momenta for different nuclei.
According to nuclear Fermi gas model, the Fermi momentum is related
to the nuclear density by $k_F=(3\pi^2\rho/2)^{1/3}$ fm$^{-1}$.
Since the nuclear density is almost a constant for very heavy nuclei,
the Fermi momentum is basically the same for very heavy nuclei.
However in terms of light nuclei, it is hard to evaluate nuclear density,
and Fermi gas model fails to give the Fermi momentum with precision.
In this analysis, we composed an empirical formula to describe
the nuclear-dependence of the Fermi momentum, which is written as
\begin{equation}
\begin{aligned}
k_F(Z,N,A)=k_F^p(1-A^{-t_1})\frac{Z}{A}+k_F^n(1-A^{-t_2})\frac{N}{A},
\end{aligned}
\label{FermiFor}
\end{equation}
where $Z$, $N$ and $A$ are proton number, neutron number and mass number respectively.
This empirical formula gives zero Fermi momenta for both the free proton
and the free neutron.
Fermi momenta of several nuclei from $^6$Li to $^{208}$Pb were
measured\cite{FermiExp-1,FermiExp-2,FermiExp-3}
by the quasielastic electron-nucleus scattering process.
The free parameters $k_F^p$, $k_F^n$, $t_1$ and $t_2$ in Eq. (\ref{FermiFor})
are determined by a fit to the experimental data.
The quality of the fit is good with $\chi^2/N_{df}=7.77/8=0.97$, which suggests that
Eq. (\ref{FermiFor}) is able to estimate the Fermi momentum with good precision.
$k_F^p$, $k_F^n$, $t_1$ and $t_2$ are obtained to be 365 MeV/c, 231 MeV/c, 0.479,
and 0.528 respectively. The average nucleon Fermi momentum of deuteron and
$^3$He are estimated to be 87 MeV/c and 134 MeV/c respectively within this approach.

\section{In-medium nucleon swelling}
\label{SecIV}

The EMC effect in the valence quark dominant region is the
most interesting subject about the nuclear modification
for both nuclear and particle physicists.
Up to date, there are so many models which can explain the principal features
of the effect. One can look at the reviews\cite{review1,review2,review3,review4}
for a good overview of the progresses on this issue.
The fact we know is that the valence quark distributions of bound
nucleon are modified by the nuclear medium environment.
Zhu and Shen\cite{ZhuShen-1,ZhuShen-2,ZhuShen-3,ZhuShen-4}
tried to calculate the nuclear medium deformation of quark distribution
functions in the constituent quark model, and achieved a big success
to well reproduce the experimental data with a few parameters.

\begin{figure}[htp]
\begin{center}
\includegraphics[width=0.41\textwidth]{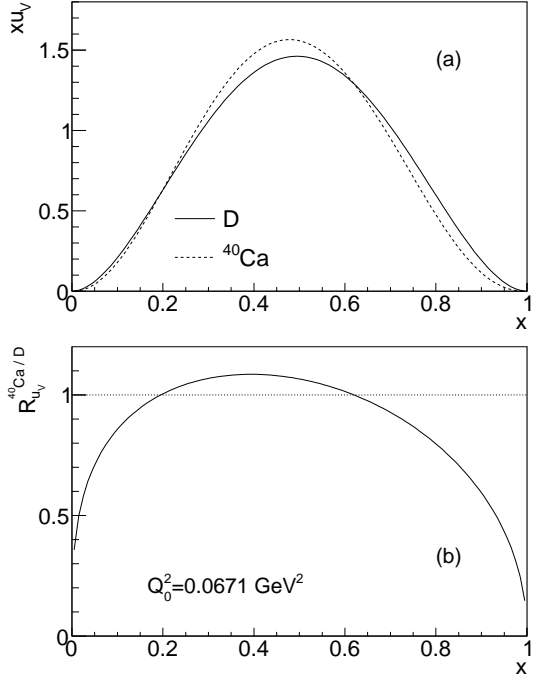}
\caption{
(a) The standard deviation of up valence quark distribution of $^{40}$Ca
is reduced compared to that of proton, due to the enlargement of the
confinement radius of in-medium nucleon; (b) Valence quark ratio of $^{40}$Ca
to deuteron at the input scale.
}
\label{SwellingEffect}
\end{center}
\end{figure}

The modification of valence quark distributions is speculated to
be related to the in-medium nucleon swelling\cite{Chen-EMC,ZhuShen-1}.
The in-medium nucleon swelling here refers to the enlargement of
the confinement scale of a valence quark, which is a basic
dynamic in strong interaction\cite{Noble1986,Arifuzzaman,Brown1988,Mardor1990,Murgia1989}.
Based on our previous work\cite{Chen-EMC,MEMPDF},
the deformations of initial valence quark distributions from the
in-medium nucleon swelling effect can be
evaluated by the Heisenberg uncertainty principle,
\begin{equation}
\begin{aligned}
\frac{\sigma(x_q^A)}{\sigma(x_q^N)}=\frac{R_N}{R_{\text{in-medium}~N}}=\frac{1}{1+\delta_A},
(q=u,d)
\end{aligned}
\label{SwellingDef}
\end{equation}
where $\delta_A$ is defined as the in-medium swelling factor,
since the spatial uncertainty of valence quark scales with the radius of the nucleon.
Here $R$ is the radius of nucleon.
$\sigma(x_q)$ in Eq. (\ref{SwellingDef}) is the standard deviation of $x$,
which is expressed as follows:
\begin{equation}
\begin{aligned}
&\sigma(x_u)=\sqrt{<x_u^2>-<x_u>^2},\\
&\sigma(x_d)=\sqrt{<x_d^2>-<x_d>^2},\\
&<x_u>=\int_0^1 x\frac{u_v(x,Q_0^2)}{2}dx,\\
&<x_d>=\int_0^1 xd_v(x,Q_0^2)dx,\\
&<x_u^2>=\int_0^1 x^2\frac{u_v(x,Q_0^2)}{2}dx,\\
&<x_d^2>=\int_0^1 x^2d_v(x,Q_0^2)dx.
\end{aligned}
\label{xDeviation}
\end{equation}
The initial valence quark distributions of free proton are well determined
so far\cite{Chen-proton,IMParton}. In this analysis, the initial
valence quark distributions of proton are taken from the Fit 2 result
of Ref. \cite{IMParton}. Under the assumption that $<x_q^A>=<x_q^N>$
and using the beta function form parametrization as $Ax^B(1-x)^C$,
the initial valence quark distributions in a nucleus
are determined by Eq. (\ref{SwellingDef}) if $\delta_A$ is known.
(See Fig. \ref{SwellingEffect}.)

The nucleon swelling factor $\delta_A$ arises from the nuclear force
and depends on the local nuclear environment, which is complicated
to calculate. Our previous work\cite{RSIE} finds that the strength
of the EMC effect linearly correlated with residual strong interaction
energy (RSIE) per nucleon. The nuclear force surely plays an important
role in the modification of confinement radius of a in-medium valence quark.
Therefore, a simple assumption is that the nucleon swelling factor
linearly scales with the RSIE, which is written as
\begin{equation}
\begin{aligned}
\delta_A=\alpha\times RSIE/A,
\end{aligned}
\label{SwellingFor}
\end{equation}
with $\alpha$ a free parameter which can be determined in the global
fit to experiments of various nuclei.

\section{Scale-dependence and gluon recombination effects}
\label{SecV}

Generally, the $Q^2$-dependence of nuclear modification factor is very weak,
which is difficult to detect. However the nuclear shadowing at small $x$
and high $Q^2$ shows clear $Q^2$-dependence for heavy nuclei\cite{NMC481.23}.
For protons, the $Q^2$-dependence of PDFs is precisely described by
DGLAP equations\cite{DGLAP-D,DGLAP-GL,DGLAP-AP}.
It was found that the $Q^2$-dependence of nuclear PDFs also obeys the leading
twist DGLAP-evolution by Eskola, Kolhinen and Ruuskanen for the first time\cite{Eskola98},
which well interprets the experimental observations.
Since then, all the global analyses on nuclear modifications by different collaborations
are performed with DGLAP equations, yet with different techniques
constructing the nuclear initial modifications or PDFs modified by various nuclear
effects including the nuclear shadowing effect. These works imply that
the collinear factorization also works for universal nPDFs.

Experimental measurements of nuclear shadowing at high $Q^2$ indicate
that the nuclear shadowing is of partonic origin, which is different
from the real photon absorption reaction. Mueller and Qiu\cite{MQ} suggest
that the shadowing in deeply inelastic scattering off nuclei is attributed to
the gluon recombination process. The parton recombination process
is the first non-trivial higher-twist correction.
The nuclear shadowing implies that the nucleons inside the nucleus
are not completely independent.
The longitudinal size of a sea quark or gluon at small $x$
can be larger than the size of a nucleon. A parton from one nucleon
which has large spatial uncertainty could leak into a neighbor and fuse
with one of partons in the neighbor nucleon.
Therefore the parton fusion process is enhanced in nuclear target.
The modifications to structure functions arising from the parton fusion
reduce the structure functions at small $x$.
The scenario of parton recombination explains the $x$-dependence of nuclear
shadowing naturally. The smaller $x$ of a parton, the bigger spatial uncertainty
the parton has, which consequence in the stronger shadowing.
Works of Qiu et al.\cite{Qiu1987,Close1989,Eskola1994,Qiu587.52,Qiu.PRL.2004}
proved that the gluon recombination process is the main
source of nuclear shadowing, which is strong.
Our previous work\cite{Chen-EMC} showed that only parton recombination process
gives a good description of the nuclear experimental data.
In this work, gluon fusion process is taken to characterize the nuclear shadowing.

DGLAP equations with GLR-MQ-ZRS\cite{GLR,MQ,Zhu-1,Zhu-2,Zhu-3} corrections are used
in this work. The gluon recombination process in GLR-MQ-ZRS corrections slows down
the gluon splitting process, resulting in lower dynamical gluon distribution in nuclei.
Therefore the dynamical sea quarks from gluon splitting are reduced, which
leads to the observed nuclear shadowing.
The DGLAP equations with gluon fusion corrections we used are written as
\begin{equation}
\begin{aligned}
Q^2\frac{dxf_{q_i^{NS}}(x,Q^2)}{dQ^2}
=\frac{\alpha_s(Q^2)}{2\pi}P_{qq}\otimes f_{q_i^{NS}},
\end{aligned}
\label{ZRS-NS}
\end{equation}
for the flavor non-singlet quark distributions,
\begin{equation}
\begin{aligned}
Q^2\frac{dxf_{\bar{q}_i^{DS}}(x,Q^2)}{dQ^2}
=\frac{\alpha_s(Q^2)}{2\pi}[P_{qq}\otimes f_{\bar{q}_i^{DS}}+P_{qg}\otimes f_g]\\
-A_{eff}\frac{\alpha_s^2(Q^2)}{4\pi\tilde{R}^2Q^2}\int_x^{1/2} \frac{dy}{y}xP_{gg\to \bar{q}}(x,y)[yf_g(y,Q^2)]^2\\
+A_{eff}\frac{\alpha_s^2(Q^2)}{4\pi\tilde{R}^2Q^2}\int_{x/2}^{x}\frac{dy}{y}xP_{gg\to \bar{q}}(x,y)[yf_g(y,Q^2)]^2,
\end{aligned}
\label{ZRS-S}
\end{equation}
for the dynamical sea quark distributions, and
\begin{equation}
\begin{aligned}
Q^2\frac{dxf_{g}(x,Q^2)}{dQ^2}
=\frac{\alpha_s(Q^2)}{2\pi}[P_{gq}\otimes \Sigma+P_{gg}\otimes f_g]\\
-A_{eff}\frac{\alpha_s^2(Q^2)}{4\pi\tilde{R}^2Q^2}\int_x^{1/2} \frac{dy}{y}xP_{gg\to g}(x,y)[yf_g(y,Q^2)]^2\\
+A_{eff}\frac{\alpha_s^2(Q^2)}{4\pi\tilde{R}^2Q^2}\int_{x/2}^{x}\frac{dy}{y}xP_{gg\to g}(x,y)[yf_g(y,Q^2)]^2,
\end{aligned}
\label{ZRS-G}
\end{equation}
for the gluon distribution, in which the factor $1/(4\pi\tilde{R}^2)$ is from the normalization of
the two-parton densities, and $\tilde{R}$ is the correlation length of the two interacting partons.
Note that the integral terms as $\int_x^{1/2}$ in above equations should be removed
when $x$ is larger than $1/2$. $\Sigma$ in Eq. (\ref{ZRS-G}) is defined as
$\Sigma(x,Q^2)\equiv\sum_jf_{q^{NS}_j}(x,Q^2)+\sum_i[f_{q^{DS}_i}(x,Q^2)+f_{\bar{q}^{DS}_i}(x,Q^2)]$.
The splitting functions of the linear terms are given by DGLAP equations, and the recombination
functions of the nonlinear terms are taken from Refs. \cite{Zhu-2,Chen-EMC,IMParton}.
The value of $\tilde{R}=3.98$ GeV$^{-1}$ is taken as the Fit 2 result in Ref. \cite{IMParton}.
$A_{eff}$ in Eq. (\ref{ZRS-S}) and (\ref{ZRS-G}) is the gluon recombination enhancement coefficient
for gluon recombination involving two nucleons in nuclear target.
Since $A_{eff}$ linearly scales with the nuclear size, a parameterized formula
is composed as
\begin{equation}
\begin{aligned}
A_{eff}=1+\beta (A^{1/3}-1),
\end{aligned}
\label{NuclRecomb}
\end{equation}
in this analysis. $(A^{1/3}-1)$ is the number of the shadowed nucleons.
$\beta$ is a free parameter, which can be determined in the global fit to nuclear data.

\section{QCD analysis}
\label{SecVI}

The initial valence distribution functions of free proton, initial
scale $Q_0$ and two-gluon correlation length $\tilde{R}$ are taken as
the Fit 2 result of our previous work\cite{IMParton}. In this analysis,
the input parton distributions at $Q_0$ of both free nucleon and nuclei
are parameterized as the beta function form as $Ax^B(1-x)^C$.
To get the initial valence quark distributions of nuclei, it is simply
just to perform some modifications by Fermi motion smearing and off-shell effect,
and in-medium nucleon swelling effect as discussed in Sec. \ref{SecIII} and \ref{SecIV}.
Since the nonperturbative input of free nucleon is already determined,
the nuclear nonperturbative input is also determined if the average Fermi momentum,
binding energy and swelling parameter $\delta_A$ are all known.
The initial valence quark distributions of nuclei are evolved to high $Q^2$
by DGLAP equations with gluon fusion process discussed in Sec. \ref{SecV}.

The running coupling constant $\alpha_s$ and the quark masses are
the same as that in our previous work\cite{IMParton}.
The fixed flavor number scheme is used to deal with heavy quarks in this work.
The suppression of dynamical strange quark is implemented to
model the flavor-dependence. Since the nuclear modification of
heavy quark is not clear in theory and their contributions are trivial
at not very high $Q^2$, only up, down and strange quarks are used
to calculate the nuclear modification factor of structure function.
In this analysis, the isospin-scalar corrected per-nucleon structure functions\cite{SLAC94}
are calculated for all nuclear targets in order to compare with the experimental data.
In theory, the isospin-scalar corrected per-nucleon nuclear structure function
is simply just $(F_2^{\text{p~in~A}}+F_2^{\text{n~in~A}})/2$
instead of $(ZF_2^{\text{p~in~A}}+NF_2^{\text{n~in~A}})/A$.

This analysis is based on the Leading Order (LO) calculations of theory.
DGLAP equations with QCD corrections of parton splitting and
parton-parton recombination are at LO, since parton-parton recombination
evolution kernels at Next-to-Leading Order are not available so far.
The running coupling $\alpha_s$ in DGLAP equations, and the Wilson
coefficient functions are also the LO results, for the consistency
of the analysis.

In this work, only two free parameters $\alpha$ and $\beta$ are used
to describe the nuclear-dependence and the $x$-dependence of nuclear modification.
The free parameters $\alpha$ and $\beta$ are determined by the least
squares method. The $\chi^2$ function is defined as
\begin{equation}
\chi^2=\Sigma_{expt.}\Sigma_{i=1}^{N_e}\frac{(D_i-T_i)^2}{\sigma_i^2},
\label{Chi2Def}
\end{equation}
where $N_e$ is the number of data points in experiment $e$, $D_i$ is a data in an experiment,
$T_i$ is the predicted value from QCD evolution, and $\sigma_i$ is the total uncertainty
combing both statistical and systematic errors.

\section{Results}
\label{SecVII}

\begin{table}[htp]
\centering
\caption{
The $\chi^2/N_{df}$ values for Fit A and Fit B.
The obtained free parameters $\alpha$ and $\beta$ of Fit A and Fit B
are also given.
}
\begin{tabular}{cccc}
\hline
~~~Fit~~~   & ~~~$\chi^2/N_{df}$~~~  & ~~~$\alpha$~~~  & ~~~$\beta$~~~ \\
\hline
A     & 1.33             & 0.00563   & 0.277   \\
B     & 1.38             & 0.00692   & 0.216   \\
\hline
\end{tabular}
\label{fitResult}
\end{table}

Two separated global fits are performed to extract the nuclear medium corrections.
One is the global fit to the data of isospin-scalar nuclei (Fit A), and the other
is the global fit to all nuclear data (Fit B). In this paper, the nuclear
corrections from Fit A and Fit B are called Set A and Set B respectively.
Since proton and neutron have obviously different structure functions,
the isospin-scalar corrections are applied for non-isospin-scalar nuclei to remove
the proton-neutron non-equality\cite{SLAC94}. The isospin-scalar corrections
strongly depend on the precise structure function information of free neutron
which have big uncertainties. The neutron structure functions at different $Q^2$
are rarely known until now with a development of the spectator tagging
technique\cite{n-structu}. The aim of Fit A is to eliminate the uncertainty of
isospin-scalar corrections, though the number of data points is cut down.
The quality of the fits is good, with both $\chi^2/N_{df}$ around 1.3,
which is shown in Table \ref{fitResult}. The $\chi^2$ for each experimental
data set are also calculated and shown in Table \ref{table_data}.
Our model by modifying the nonperturbative input with different nuclear effects
agrees well with the experimental data. The parameters $\alpha$ and $\beta$ are
obtained and shown in Table \ref{fitResult}, which are used to describe the nuclear
dependence of  nucleon swelling and shadowing respectively.

\begin{figure}[htp]
\begin{center}
\includegraphics[width=0.5\textwidth]{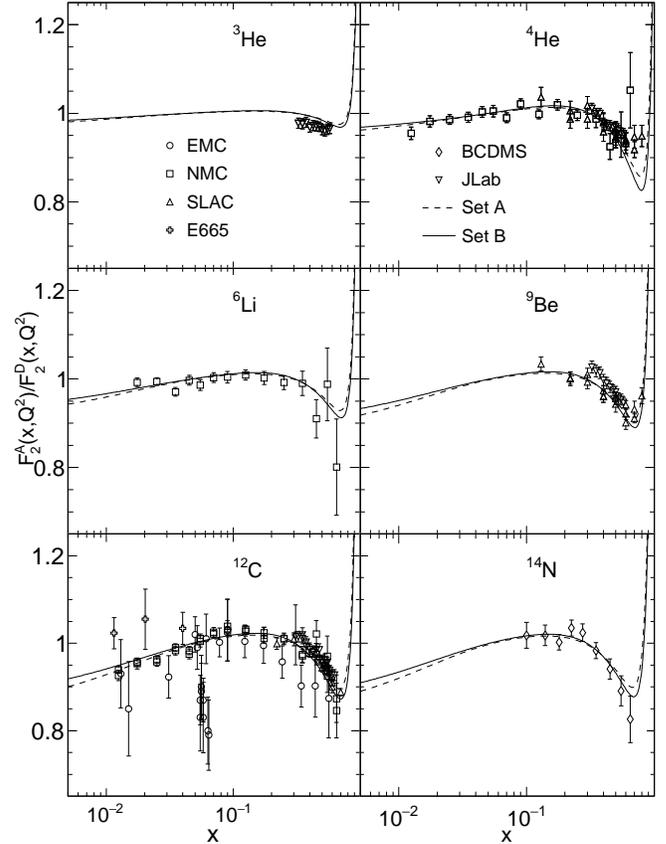}
\caption{
The global fit results of structure function ratios are compared to experimental
data from EMC\cite{EMC202.603,EMC211.493,EMC333.1}, NMC\cite{NMC441.3,NMC441.12},
SLAC\cite{SLAC94}, E665\cite{E66595}, BCDMS\cite{BCDMS85}, and JLab\cite{JLab09}
for light nuclei targets.
The $Q^2$ of both Set A and Set B data are 5 GeV$^2$ in this figure.
}
\label{LightNuclei}
\end{center}
\end{figure}

\begin{figure}[htp]
\begin{center}
\includegraphics[width=0.5\textwidth]{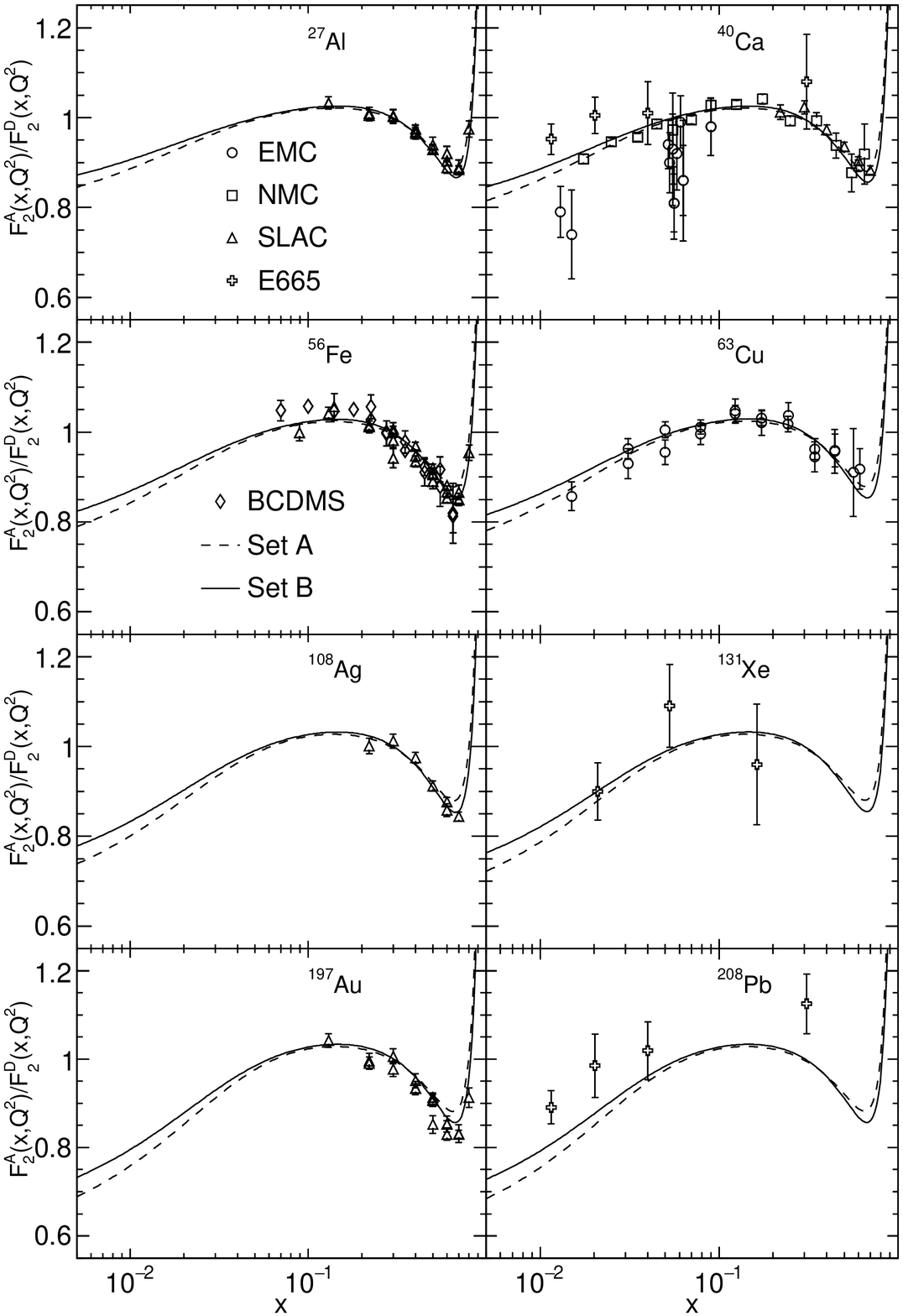}
\caption{
The global fit results of structure function ratios are compared to experimental
data from EMC\cite{EMC211.493,EMC333.1,EMC57.211}, NMC\cite{NMC441.3},
SLAC\cite{SLAC94}, E665\cite{E66592,E66595}, and BCDMS\cite{BCDMS85,BCDMS87}
for heavy nuclei targets.
The $Q^2$ of both Set A and Set B data are 5 GeV$^2$ in this figure.
}
\label{HeavyNuclei}
\end{center}
\end{figure}

\begin{figure}[htp]
\begin{center}
\includegraphics[width=0.5\textwidth]{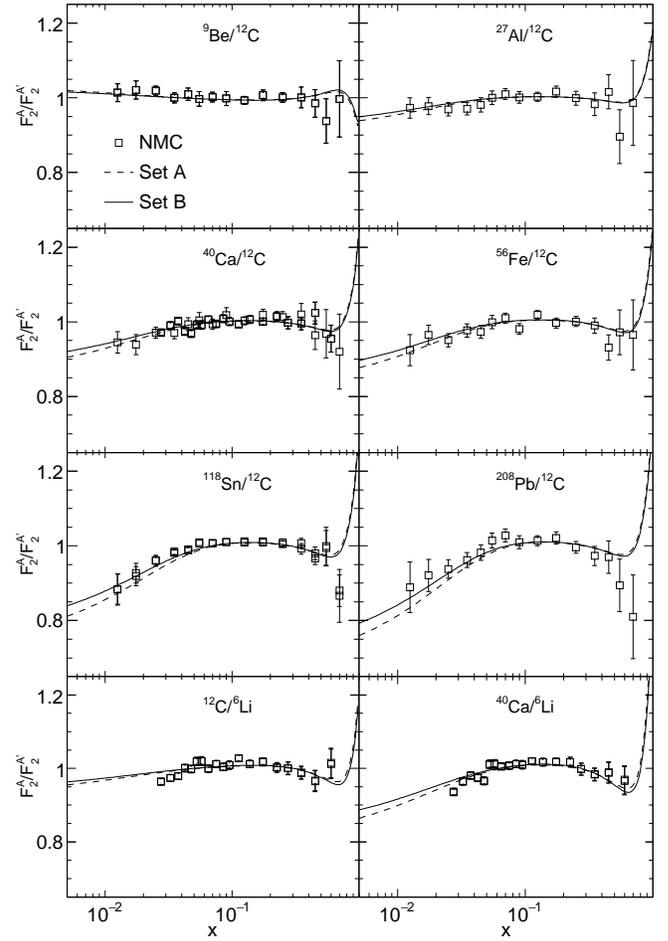}
\caption{
The global fit results of structure function ratios between two
different nuclei targets are compared to the precise NMC
data\cite{NMC53.73,NMC441.3,NMC481.3,NMC481.23}.
The $Q^2$ of both Set A and Set B data are 5 GeV$^2$ in this figure.
}
\label{HeavyNuclei_NMC}
\end{center}
\end{figure}

The global fit results compared with the experimental measurements are shown in
Figs. \ref{LightNuclei}, \ref{HeavyNuclei} and \ref{HeavyNuclei_NMC}.
The obtained data sets are consistent with the experimental data from light nuclei
(Fig. \ref{LightNuclei}) target to heavy nuclei target (Fig. \ref{HeavyNuclei}).
The most precise nuclear shadowing data were measured by NMC Collaboration
using two solid target sets\cite{NMC481.3}. Fig. \ref{HeavyNuclei_NMC} shows the
result of Fit A and Fit B compared to the precise NMC data of per-nucleon
structure function ratios of one nucleus to another nucleus. It is also clearly
demonstrated that the nuclear dependence and $x$-dependence are well reproduced
in the global analysis in these figures. The simple formulas expressed as
Eqs. (\ref{SwellingFor}) and (\ref{NuclRecomb}) are good approximations
to model the nuclear dependence of nuclear modifications.
The $x$-dependence of nuclear shadowing is well reproduced by the parton-parton
recombination corrections.

\begin{figure}[htp]
\begin{center}
\includegraphics[width=0.43\textwidth]{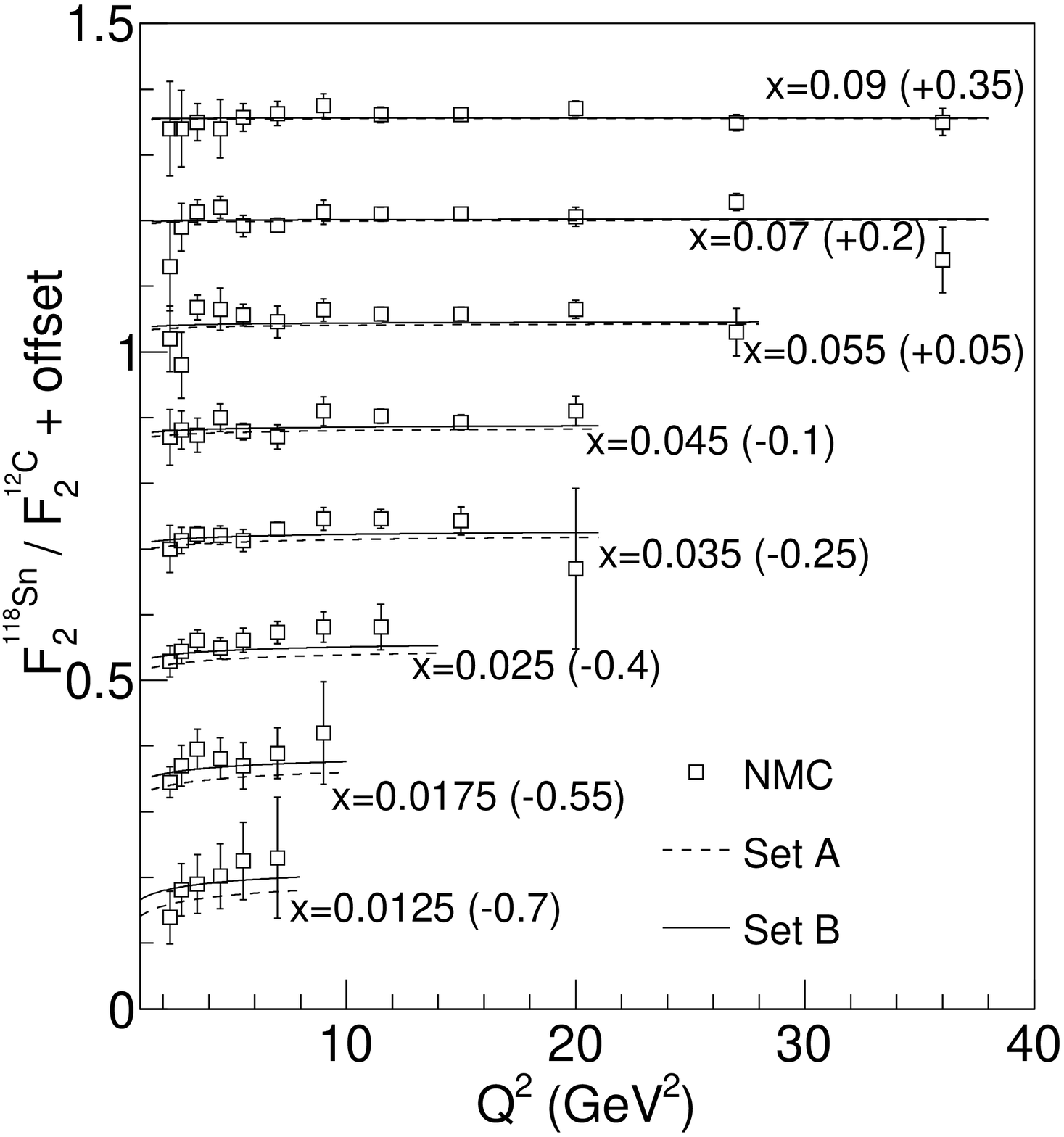}
\caption{
The $Q^2$-dependence of the structure function ratio of $^{118}$Sn to
$^{12}$C at different $x$. The squares are the NMC data\cite{NMC481.23},
and the curves are the global fit results.
The values in the brackets are the offsets added to the structure function ratios.
}
\label{Q2_Depen}
\end{center}
\end{figure}

Fig. \ref{Q2_Depen} shows the $Q^2$-dependence of the nuclear modification.
The predictions by the DGLAP equations with nonlinear corrections are consistent
with the NMC data of $^{118}$Sn. The scale-dependence of structure function ratio
is very weak, because the parton evolution at high $Q^2$ is almost the same for
two different nuclei. In our approach, the only difference of evolutions for
different nuclei comes from the higher-twist corrections to DGLAP equations.
So far the $Q^2$-dependence of nuclear shadowing around $x=0.01$ is in agreement
with the DGLAP equations with parton-parton recombination corrections, which suggest
that the DGLAP equations are universal for hadron systems. More experimental
data at small $x$ with wide kinematic range are needed to further check the
DGLAP evolution of nPDFs.

\begin{figure}[htp]
\begin{center}
\includegraphics[width=0.41\textwidth]{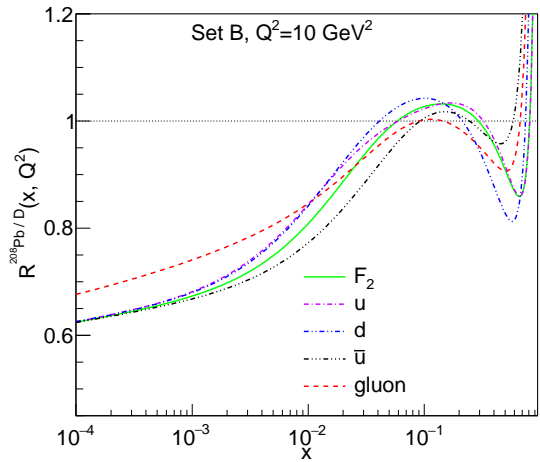}
\caption{
The nuclear modification factors of different flavors of $^{208}$Pb predicted
by this work. The flavor-dependence of nuclear modification factors are shown weak.
}
\label{Flavor_Depen}
\end{center}
\end{figure}

Fig. \ref{Flavor_Depen} shows the predicted flavor-dependence of nuclear modifications
on parton distributions of this work. In this analysis, no strong flavor-dependence
of nuclear modifications is shown. Nevertheless, the nuclear modification of gluon
shows some difference. The shadowing of gluon is a little weaker than that of sea quarks,
and its anti-shadowing is quite small. In small $x$ region ($x<10^{-3}$), the strengths
of shadowing effect of both quark and anti-quark are the same. However, the shadowing
magnitude of quark distribution around $x=0.01$ is weaker than that of anti-quark
distribution, which is due to the weak shadowing effect of valence quark distribution
contributing to the shadowing of quark distribution. For the EMC effect, we find that
the magnitude of the EMC effect of gluon is smaller than that of valence quarks,
and the magnitudes of the EMC effect of sea quarks are smaller than those of gluon.
In our approach, the nucleon swelling effect is applied on the nonperturbative input
which merely consists of valence quarks. Therefore the gluons generated by the
radiation of valence quarks have weaker EMC effect, and the sea quarks from the gluon
splitting have even weaker EMC effect. In the semimicroscopic model by KP, the sea
quarks also have very small EMC effect\cite{KP-DY}. The valence quarks also have the strongest
anti-shadowing effect, and, the anti-shadowing effect of sea quarks is stronger
than that of gluon.
The dynamical sea quarks are generated from the gluon splitting in the QCD evolution.
Therefore the modification factors of $\bar{u}$, $\bar{d}$, and $\bar{s}$ are the same.

\begin{figure}[htp]
\begin{center}
\includegraphics[width=0.48\textwidth]{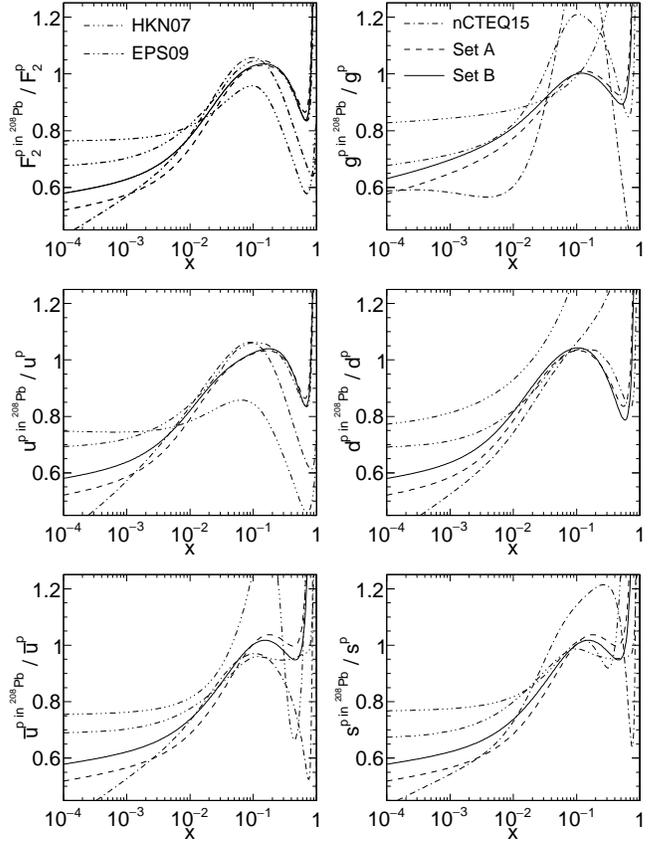}
\caption{
The nuclear modification factors of different flavors are shown
with other widely used nPDFs, such as EPS09\cite{EPS09}, nCTEQ15\cite{nCTEQ15}, and HKN07\cite{HKN07}.
The resolution scales for these nuclear modifications are all at $Q^2=5$ GeV$^2$.
}
\label{NPDF_Comp}
\end{center}
\end{figure}

Fig. \ref{NPDF_Comp} shows the predicted nuclear modification factors of different
flavors for $^{208}$Pb target, compared with other widely used nPDFs.
There show a lot of difference among nPDFs from different groups
in terms of the nuclear modification factors.
For the structure function ratio, our result is very close to the prediction of EPS09,
except the shadowing effect. The structure function shadowing of this analysis at small
$x$ is a little stronger than that of EPS09 analysis, and it is close to the result of
nCTEQ15 around $x\lesssim 0.01$. For the gluon shadowing effect, this analysis gives
similar prediction as that of EPS09 and nCTEQ15 around $x\sim 10^{-4}$.
However, for anti-shadowing effect of gluon distribution, the prediction of this work is
special, which has very weak anti-shadowing. The predictions of HKN07, EPS09, and nCTEQ15
all have very big anti-shadowing for gluon distribution. This is an interesting
window to test the nuclear dynamical gluon model of this analysis. For the shadowing effect
of sea quarks, our predictions are stronger than those of HKN07 and EPS09, but they are weaker
than predictions by nCTEQ15. In terms of the EMC effect for valence quark
and sea quark distributions, the predictions of this analysis are close to that of EPS09.
The EMC effects of $^{208}$Pb predicted by HKN07 and nCTEQ15 are stronger than our result.
More nuclear DIS data, especially in small $x$ region ($x\lesssim 10^{-3}$)
are needed to distinguish the predictions by different nPDFs analyses.

\begin{figure}[htp]
\begin{center}
\includegraphics[width=0.41\textwidth]{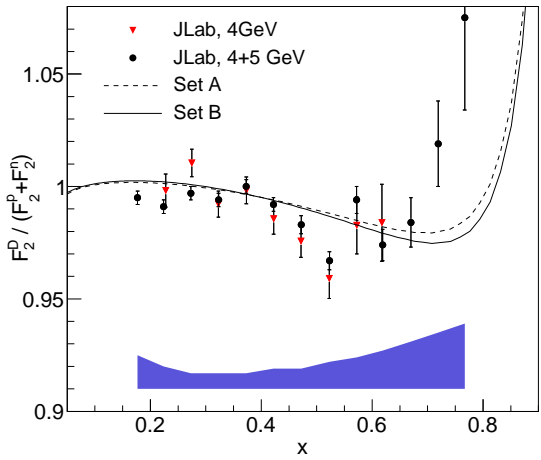}
\caption{
The predicted nuclear correction of deuteron compared with the state-of-art
measurement at JLab\cite{d-EMC-ref}. The $Q^2$ of both Set A and Set B data
are 2 GeV$^2$ in this figure.
The JLab data is in the range of $1$ GeV$^2$ $<Q^2<$ 4.8 GeV$^2$.
}
\label{D_EMC}
\end{center}
\end{figure}

With the obtained parameters $\alpha$ and $\beta$, the nuclear correction of deuteron
is predicted under the framework of Eq. (\ref{SwellingFor}) and (\ref{NuclRecomb}).
Although deuteron is a very loosely bound nucleus, its structure function is slightly
different from that of free proton and free neutron. The state-of-art measurement
of the structure function ratio $R_{EMC}^d=F_2^d/(F_2^n+F_2^p)$ is performed
at JLab\cite{d-EMC-ref} in the kinematic region of $W>1.4$ GeV and $Q^2>1$ GeV$^2$.
Fig. \ref{D_EMC} shows the comparison of the predicted nuclear correction of deuteron
with the JLab data. In the EMC effect region, the prediction of the global analysis
is in good agreement with experimental data. The predicted Fermi motion smearing effect
is weaker than the measured data. Fermi momentum of 87 MeV for deuteron used
in the calculation maybe is small. The realistic deuteron wave functions are needed,
or we need a more complicated formula for Fermi smearing.

\section{nIMParton package}
\label{SecVIII}

We provide a C++ library named nIMParton to access the obtained nuclear modification
factors of various nuclei for practical applications, so as to avoid the complicated
nuclear effect calculations and the slow QCD evolution with parton-parton recombination
corrections. The C++ package is now available from us via email,
the WWW\cite{nIMParton}, or downloading by the git command\cite{gitnIMParton}.
Two data sets of the global analysis result, called data set A
and data set B, are provided by the package, which is
discussed at length in Sec. \ref{SecVII}. The given nuclear modification factor
by nIMParton is the result from a interpolation of the grid table calculated
by the models.

The package consists of a C++ class named nIMParton which gives the
interface to the nPDFs. The constructor function nIMParton(Z, A)
is used to choose a nuclear target. nIMParton has two important methods
getRToN(Iparton, X, Q2) and  getRToD(Iparton, X, Q2), which are used to get the
parton distribution ratios of a nucleus to free nucleon and deuteron respectively,
and suggested to be called in users' programs.
Iparton set as -3, -2, -1, 0, 1, 2, and 3 corresponds to getting parton distribution
ratios $R_{\bar{s}}$, $R_{\bar{d}}$, $R_{\bar{u}}$, $R_g$, $R_{u}$, $R_{d}$, and
$R_{s}$ respectively. Another important method of nIMParton is setDataSet.
setDataSet(1) corresponds to use the data set A,
and setDataSet(2) corresponds to use the data set B.
For isospin-scalar nuclei, the data set A is recommended.
Nuclear modifications on heavy quarks are not determined in this work.
We suggest using $R_c=R_b=R_g$ if they are needed, since heavy quarks are mainly
produced by gluons. This is a feasible solution, as this analysis shows
relatively weak flavor-dependence of nuclear modifications.

The nIMParton package gives only the nuclear modification factors which are
calculated as the parton distribution ratios of a bound nucleon to a free nucleon.
Under the assumption that the nuclear modification factors of parton
distributions are the same for both the bound proton and the bound neutron,
the nPDFs can be constructed by the following formula,
\begin{equation}
\begin{aligned}
f_i^A(x,Q^2) =\\
\frac{ZR_i^{A/N}(x,Q^2)f_i^p(x,Q^2)+(A-Z)R_i^{A/N}(x,Q^2)f_i^n(x,Q^2)}{A},\\
\end{aligned}
\label{NPDF_Cal}
\end{equation}
where $i$, $A$ and $Z$ are the flavor index, the mass number and proton number
of a nucleus respectively. The proton PDFs $f_i^p$ are precisely determined
by many collaborations through decades of development, and the neutron PDFs can
easily be given by the isospin symmetry. For modern PDFs of free proton,
we can look at the work\cite{IMParton} and the references therein.

\section{Discussions and summary}
\label{SecIX}

Two data sets of nuclear modification factors are given by a global
analysis of nuclear DIS data worldwide. Data set A is from the global fit
to the data of only isospin-scalar nuclei, and data set B is from the global
fit to all nuclear data.
Both data sets are in excellent agreement with the measured structure function
ratios. The small difference between set A and set B is on the shadowing in the small
$x$ region ($x<0.01$). The predicted nuclear correction of deuteron
 is consistent with the state-of-art
measurement at JLab (not included in the global fit).
Comparisons with other nPDFs are shown, such as HKN07,
EPS09, and nCTEQ15. Nuclear modification factors of parton distributions
from different nPDFs analyses show many differences, especially at small $x$.
More nuclear data at small $x$ are needed to improve the precision of nuclear
corrections at small $x$. Our prediction
of the strength of quark shadowing is weaker than the prediction by nCTEQ15, and
stronger than predictions by EPS09 and HKN07. On the shadowing of gluon
distribution at small $x$, the prediction of this work is close to that
of EPS09 and nCTEQ15. The interesting characteristic of this analysis is the
gluon anti-shadowing. The gluon ratio of a heavy nucleus to deuteron
is almost one in the anti-shadowing region around $x=0.1$.
KT analysis also gives no anti-shadowing of gluon distribution,
but it also gives no noticeable gluon shadowing\cite{KT16}.
The obtained data sets are expected to be the good options as the nPDFs input
for d-A or A-A collisions. A C++ package is provided to interface
with the obtained nuclear modification factors. The nPDFs are constructed
by Eq. (\ref{NPDF_Cal}).

There are a lot of free parameters to describe the $x$-dependence and
nuclear dependence of nuclear modification for the common global analyses
of nPDFs. Same as the KP analysis\cite{KP06,KP07}, the global analysis are
based on a combination of different nuclear models in this paper. By this method,
the number of free parameters to describe the nuclear modifications
is reduced to only two. Based on the dynamical parton model and models
of nuclear effects, the method eliminates
the arbitrary of parametrization and of the solution of the optimal fit
in high dimensional space of parameters. The nuclear modification factors
are more reliable, since they have more theoretical constraints.

The nuclear dependence for nuclear modifications in this work is very
predictive, which will be tested by measurements of unmeasured nuclei
at future EIC (Electron-Ion Collider) project. The $Q^{2}$-dependence is weak.
However the observed weak $Q^{2}$-dependence at small $x$ and low $Q^2$ support
the description of the DGLAP equations with parton-parton recombination corrections.
The flavor-dependence of nuclear modification factors in this analysis is not big.
The gluon shadowing is just a little weaker than that of quarks.
The anti-shadowing effect of valence quarks
is stronger than that of sea quarks, and the anti-shadowing effect of sea quarks
is stronger than that of gluons.
The magnitude of the EMC effect of valence quarks (or quarks at large $x$)
is stronger than that of gluon, and the magnitude of the EMC effect of gluon
is stronger than that of sea quarks.
This flavor-dependence is sensitive to test
the nuclear models used in this analysis.

The nuclear modification of gluon distribution can be fixed in the global analysis
under the dynamical parton model using combined nuclear models.
The obtained nuclear gluon distribution is
expected to be predictive and more reliable. The nuclear gluon distribution is vital for
the vector boson, jet or other hadron productions at high energy. One window to test
the nuclear modification of gluon distribution in this work is the $J/\Psi$ production
in d-A or A-A collision. We will discuss it elsewhere.

\noindent{\bf Acknowledgments}:
We thank Wei Zhu, Pengming Zhang, Jianhong Ruan and Fan Wang
for the helpful and fruitful discussions. This work was supported by
the National Basic Research Program (973 Program Grant No. 2014CB845406),
and Century Program of Chinese Academy of Sciences Y101020BR0.

\end{document}